\documentclass{ws-procs975x65}

\usepackage{amsmath}
\usepackage{amsfonts}
\usepackage{amssymb}

\begin{document}

\title{Nieh-Yan Invariant and Fermions in Ashtekar-Barbero-Immirzi Formalism}

\author{Simone Mercuri}

\address{Dipartimento di Fisica, Universit\`a di Roma ``La Sapienza'', 
Piazzale Aldo Moro 5, I-00185, Rome, Italy}
\address{ICRA --- International Center for Relativistic Astrophysics}

\maketitle

\abstracts{In order to introduce an interaction between gravity and fermions in the Ashtekar-Barbero-Immirzi formalism without affecting classical dynamics a non-minimal term is necessary. The non-minimal term together with the Holst modification to the Hilbert-Palatini action reconstruct the Nieh-Yan invariant. As a consequence the Immirzi parameter, differently from the minimal coupling approach, does not affect the classical dynamics, which is described by the Einstein-Cartan action.}

The introduction by Ashtekar of self-dual $SL(2,\mathbb{C})$ connections\cite{ash86}, which reduces the phase space of General Relativity to that of a Yang-Mills gauge theory, has given a boost to the program of a background independent quantum theory of gravity and has finally led to the formulation of the so called \emph{Loop Quantum Gravity}\cite{rov04,ashlew04}. The use of the complex Ashtekar connections simplifies remarkably the Hamiltonian constraints of the theory, which are reduced to a polynomial form, but, on the other hand, in order to assure that the evolution be real, a reality condition is necessary. Implementing the reality condition at the quantum level is a very difficult task, so the real Barbero connections\cite{bar95} are in general preferred, even though the Hamiltonian scalar constraint results more complicate and non-polynomial. The relation existing between the complex Ashtekar connections and the Barbero's real ones was clarified by Immirzi\cite{imm97-1}, with the introduction of the so called Immirzi parameter $\beta$, in the definition of the new connections. Being introduced via a canonical transformation the Immirzi parameter does not affect the classical dynamics, but it has important effects in the quantum non-perturbative regimes as explained in\cite{rovthi98}. This double role of the parameter $\beta$ suggests an analogy with the parameter $\theta$ in QCD\cite{gamobrpul98}. In fact the analogy exists, because both the parameters results to be multiplicative factors in front of topological terms(\footnote{It is worth noting that the adjective topological is generally referred to objects like the integrals of Pontryagin or Chern classes, which, if the space is compact, depend only on the topological characteristics of the manifold, but it is often, even though improperly, used referring to the object multiplying the Immirzi parameter, which does not belong neither to the Chern nor to the Pontryagin classes and is defined on a pseudo-Riemannian manifold.}), as the Holst covariant approach clearly shows\cite{hol96}. Basically the Holst action contains a modification with respect to the Hilbert-Palatini action, which vanishes once the torsionless second Cartan structure equation is satisfied, if torsion is present things could change. As a consequence spinor fields could affect this picture. In fact, as well known, the presence of spinors in the dynamics generates a non-vanishing torsion 2-form, which modifies the Cartan structure equation and, in the usual Einstein-Cartan theory, yields a Fermi-like four spinors interaction term; the questions we want to address in this brief paper are: Does the Holst modification to the Hilbert-Palatini action affect the Einstein-Cartan picture? And then: If it is the case, is it possible to postulate a non-minimal coupling in order the resulting effective theory is the Einstein-Cartan one? Does the non-minimal coupling any geometrical meaning?

The answer to the first question is addressed in a couple of papers and confirms what initially expected, in fact, minimally coupling spinors to the gravitational field described by the Holst action and variating the total action with respect to the Lorentz valued connection, one finds a non-vanishing right side in the Cartan structure equation. After having extracted the expression of the right-hand side 2-form:
\begin{equation}\label{torsion}
T^{a}=\,-\frac{1}{4}\,\frac{\beta^2}{1+\beta^2}\left(\epsilon^{a b}_{\phantom1\phantom2 c d}+\frac{1}{\beta}\,\delta^{[a}_{c}\delta^{b]}_{d}\right)J_{(A)\,b}e^{c}\wedge e^{d},
\end{equation}
(where $J_{(A)}^a=\overline{\psi}\gamma^a\gamma^5\psi$) one immediately realizes it differs from the Torsion tensor coming out in the Einstein-Cartan theory, both for the presence of an additional term and for the dependence on the Immirzi parameter (obviously as soon as the limit $\beta\rightarrow\infty$ is calculated the 2-form above reduces to the torsion of the Einstein-Cartan theory): as a consequence also the effective action depends on the Immirzi parameter\cite{perrov05,fremintak05}. It is worth noting that the 2-form in line (\ref{torsion}) cannot be associated with the torsion of space-time, even though it represents the right hand side of a dynamical equation analogue to the structure equation of the Einstein-Cartan theory. The point is that the 2-form (\ref{torsion}) contains a pseudo-vectorial term, which  cannot be traced back to anyone of the irreducible components of the torsion tensor\cite{RovMer07-discussion} (\footnote{We stress that, even though the resulting connection contains two parts with different transformation properties under the sector $P$ of the Lorentz group, the effective theory does not violate the parity discrete symmetry.}).

The resulting modification to the Einstein-Cartan effective action and the classical role the Immirzi parameter would play in this framework, suggest to search for a different formulation of the interaction between gravitational and spinor fields. In particular, we found that using the following non-minimal action
\begin{align}\label{new action}
\nonumber S\left(e,\omega,\psi,\overline{\psi}\right)=&\,\frac{1}{2}\int\left(\frac{1}{2}\,\epsilon_{a b c d}\,e^{a}\wedge e^{b}\wedge R^{c d}-\frac{1}{\beta}\,e_{a}\wedge e_{b}\wedge R^{a b}\right)
\\
&+\frac{i}{2}\int\star\; e_a\wedge\left[\overline{\psi}\gamma^a\left(1-\frac{i}{\beta}\gamma_5\right) \mathcal{D}\psi-\overline{\mathcal{D}\psi}\left(1-\frac{i}{\beta}\gamma_5\right)\gamma^a\psi\right],
\end{align}
we can describe the interaction between the gravitational field and spinor matter without affecting the effective limit and leading to a natural generalization of the Holst approach\cite{mer06}.
In fact, the above action reduces to the usual Einstein-Cartan effective action once the second Cartan structure equation is satisfied and generates consistent dynamical equations for every value of the Immirzi parameter(\footnote{It is worth noting that the minimal approach previously described applies only to real values of the Immirzi parameter.}), generalizing the Ashtekar-Romano-Tate one\cite{mer06}. The non-minimal spinor coupling term together with the Holst modification reconstruct, once the Cartan structure equation is satisfied, the so called Nieh-Yan invariant\cite{nieyan82}. In other words we have(\footnote{For the details of the demonstration and a brief discussion of the Nieh-Yan topological term we address the reader to\cite{mer06}.})
\begin{align}\label{equality II}
\frac{1}{2\beta}\int\left[e_{a}\wedge e_{b}\wedge R^{a b}+\star\; e_a\wedge\left(\overline{\psi}\gamma_5\gamma^a\mathcal{D}\psi-\overline{\mathcal{D}\psi}\gamma^a\gamma_5\psi\right)\right]=\frac{1}{2\beta}\int d\left(T_a\wedge e^{a}\right).
\end{align}
Moreover the non-minimal spinor action (\ref{new action}) can be, unexpectedly, separated in two independent actions with different weights depending on the Immirzi parameter, where the respective interaction terms contain the self-dual and anti-self-dual Ashtekar connections; this suggests to search for a similar separation in the Holst action, in order to rewrite the total action as the sum of two actions describing independently the self-dual and anti-self-dual sector of the complete theory. This separation is in fact possible and, as noted by Alexandrov in\cite{ale06}, referring to the pure gravitational case, both the constraints and the reality condition simplify using the self-dual and anti self-dual Ashtekar connections as separate variables. On the other hand, once one realizes that the real Barbero connections can be written as a weighted sum of self-dual and anti-self-dual connections with weights depending on the Immirzi parameter, the calculation of the Hamiltonian constraints for the real connections can be performed starting, directly, from the separated action.

\end{document}